\documentclass[preprintnumbers,article,amsmath,amssymb,floatfix,10pt,prd,twocolumn,
superscriptaddress,nofootinbib]{revtex4-2}
\usepackage{bm}
\usepackage{amsfonts}
\usepackage{latexsym}
\usepackage[latin1]{inputenc}
\usepackage{graphicx}
\usepackage{amsmath}
\usepackage{palatino}
\usepackage{mathpazo}
\usepackage{textcomp}
\usepackage{float}
\usepackage{booktabs}
\usepackage{dcolumn}
\usepackage{ragged2e}
\usepackage{hyperref}
\hypersetup{colorlinks,citecolor=blue}
\hypersetup{colorlinks=true,linkcolor=red,filecolor=magenta,    urlcolor=blue}
\usepackage{amsmath}
\usepackage{xcolor}
\usepackage{orcidlink}
\usepackage{epsfig}
\usepackage{caption}
\usepackage{subcaption}
\usepackage{commath}
\def\jnl@style{\it}
\def\aaref@jnl#1{{\jnl@style#1}}

\def\aaref@jnl#1{{\jnl@style#1}}

\def\aj{\aaref@jnl{AJ}}                   
\def\apj{\aaref@jnl{ApJ}}                 
\def\apjl{\aaref@jnl{ApJ}}                
\def\apjs{\aaref@jnl{ApJS}}               
\def\apss{\aaref@jnl{Ap\&SS}}             
\def\aap{\aaref@jnl{A\&A}}                
\def\aapr{\aaref@jnl{A\&A~Rev.}}          
\def\aaps{\aaref@jnl{A\&AS}}              
\def\mnras{\aaref@jnl{Mon.~Not.~Roy.~Astron.~Soc.}}             
\def\prd{\aaref@jnl{Phys.~Rev.~D}}        
\def\prc{\aaref@jnl{Phys.~Rev.~C}}  
\def\prl{\aaref@jnl{Phys.~Rev.~Lett.}}    
\def\qjras{\aaref@jnl{QJRAS}}             
\def\skytel{\aaref@jnl{S\&T}}             
\def\ssr{\aaref@jnl{Space~Sci.~Rev.}}     
\def\zap{\aaref@jnl{ZAp}}                 
\def\nat{\aaref@jnl{Nature}}              
\def\aplett{\aaref@jnl{Astrophys.~Lett.}} 
\def\apspr{\aaref@jnl{Astrophys.~Space~Phys.~Res.}} 
\def\physrep{\aaref@jnl{Phys.~Rep.}}      
\def\physscr{\aaref@jnl{Phys.~Scr}}       
\def\commat{\aaref@jnl{Comm.~Math.~Phys.}}              
\def\science{\aaref@jnl{Science}}               
\def\cqg{\aaref@jnl{Classical Quant.~Grav.}}            
\def\jpcs{\aaref@jnl{JPCS}}                                     
\def\ijmpd{\aaref@jnl{Int.~J.~Mod.~Phys.~D}}                    
\def\grg{\aaref@jnl{Gen.~Relat.~Gravit.}}               
\def\rpp{\aaref@jnl{Rep.~Prog.~Phys.}}          
\def\npa{\aaref@jnl{Nucl.~Phys.~A}}        
\def\lrr{\aaref@jnl{Living Rev.~Rel.}}                   
\def\jcap{\aaref@jnl{J.~Cosmology Astropart.~Phys.}}    
\def\rmp{\aaref@jnl{Rev.~Mod.~Phys.}}   
\def\epjc{\aaref@jnl{Eur.~Phys.~J.~C}} 
\def\plb{\aaref@jnl{~Phy.~Lett.~B}} 
\def\mpla{\aaref@jnl{Mod.~Phy.~Lett.~A}} 
\def\arxiv{\aaref@jnl{arxiv.org}}

\allowdisplaybreaks[1]

\begin{document}
\color{black}       
%
\title{Statistical and Observation Comparison of Weyl-Type $f(Q,T)$ Models with the $\Lambda$CDM Paradigm}
\author{Gaurav N. Gadbail\orcidlink{0000-0003-0684-9702}}
\email{gauravgadbail6@gmail.com}
\affiliation{Department of Mathematics, Birla Institute of Technology and
Science-Pilani, Hyderabad Campus, Hyderabad-500078, India.}
\author{Himanshu Chaudhary\orcidlink{0000-0002-6376-0707}}
\email{himanshu.chaudhary@ubbcluj.ro}
\affiliation{Department of Physics, Babes-Bolyai University, Kogalniceanu Street, Cluj-Napoca, 400084, Romania,}
\affiliation{Research Center of Astrophysics and Cosmology, Khazar University, Baku, 41 Mehseti Street, AZ1096, Azerbaijan}
\author{Amine Bouali\orcidlink{0000-0002-6376-0707}}
\email{a1.bouali@ump.ac.ma}
\affiliation{Laboratory of Physics of Matter and Radiation, Mohammed I University, BP 717, Oujda, Morocco,}
\affiliation{Higher School of Education and Training, Mohammed I University, BP 717, Oujda, Morocco.}
\author{P.K. Sahoo\orcidlink{0000-0003-2130-8832}}
\email{pksahoo@hyderabad.bits-pilani.ac.in}
\affiliation{Department of Mathematics, Birla Institute of Technology and
Science-Pilani, Hyderabad Campus, Hyderabad-500078, India.}

\begin{abstract}
We study the $f(Q,T)$ gravity in the framework of Weyl geometry (known as Weyl-type $f(Q,T)$ gravity), where $Q$ denotes the non-metricity scalar, and $T$ denotes the energy-momentum tensor trace. In this work, we consider the $f(Q,T)$ model, which is defined as $f(Q,T)=\alpha Q^{m+1}+\frac{\beta}{6\kappa^2}T$ and investigating two scenarios: $(I)$ $m=0$ (linear model) and $(II)$ $m\neq 0$ (nonlinear model). For both scenarios, we find the explicit solution for the field equations by using the barotropic equation of state as $p=w\rho$, where $w$ is the equation-of-state (EoS) parameter. Further, we study the obtained solutions statistically using the $Pantheon^+$ (Without SHOES Calibrated) dataset with 1701 data points. For both models, the best-fit values of model parameters for $1-\sigma$ and $2-\sigma$ confidence level.  The higher Hubble constant values in both models emphasize the presence of Tension. We statistically compare our models to the $\Lambda$CDM model using ${{\protect\chi}^2_{min}}$, ${{\protect\chi}^2_{red}}$, $AIC$, $\Delta AIC$, $BIC$ and $\Delta BIC$. We also examine cosmological parameters such as deceleration and EoS parameters to determine the current acceleration expansion of the Universe. Furthermore, we test our model using $Om$ diagnostic and compare it to the $\Lambda$CDM model to determine its dark energy profile. Finally, we draw the conclusion that statistically speaking, both linear and nonlinear models show good compatibility with the $\Lambda$CDM model.
\end{abstract}

\maketitle

\date{\today}

\section{Introduction}
Modern astronomical discoveries, such as high redshift Type Ia supernovae (SNe Ia), observational Hubble parameter data (OHD), matter power spectra, cosmic microwave background radiation (CMBR), and others \cite{1:2001qse,2,3,4,5,6} have demonstrated that our Universe is currently entered an accelerated expansion era. To explain this Universe's acceleration mechanism, the cosmologist suggested the presence of a new negative pressure fluid called Dark energy (DE).
The most well-known candidate for DE is the so-called $\Lambda$-cold-dark-matter ($\Lambda$CDM), which has been confirmed to be very prosperous in explaining many aspects of the observed Universe. Despite its enormous success, the $\Lambda$CDM model has some issues, such as cosmic coincidence and fine-tuning problems \cite{7,8}.  As a result, this prompts the investigation of potential alternatives. In this view, there are two types of alternative possibilities: modifying matter content in the Einstein-Hilbert action and modifying the gravitational part of the Einstein-Hilbert action called modified gravity. The modified theory has recently emerged as one of the most favored alternatives for explaining the DE problem. In order to explain the acceleration of the Universe in early and late times, several authors have discussed various modified theories such as $f(R)$ theory (where $R$ is Ricci scalar) \cite{9,10,clifton2012modified,nojiri2011unified,nojiri2017modified,Odintsov:2023weg}, $f(\mathcal{T})$ theory (where $\mathcal{T}$ is Torsion scalar) \cite{11}, $f(R,T)$ theory (where $T$ denotes the energy-momentum tensor trace) \cite{12}, $f(\mathcal{T},T)$ theory \cite{harko2014f}, $f(G)$ theory (where $G$ is Gauss-Bonnet invariant) \cite{13}, and $f(G, T)$ theory \cite{14}, etc. \\\\
Recently, a new theory of gravity (known as symmetric teleparallel theory) has been developed based on a geometrical variable called non-metricity $Q$, which describes the properties of gravitational interaction \cite{15}. Geometrically, the non-metricity $Q$ characterizes a vector's length interpretation under parallel transport. Further, the symmetric teleparallel theory was developed into the $f(Q)$ theory \cite{16}. Various physical and geometrical aspects of $f(Q)$ gravity theory have been studied in the literature \cite{17,18,19,20,22,23,24,Gadbail_APJ_2024}. Furthermore, the $f(Q)$ theory of gravity was investigated through observational data in \cite{21,anagnostopoulos2021first} . In order to extract constraints on $f(Q)$ gravity, Anagnostopoulos et al. utilize the BBN formalism and observations in \cite{anagnostopoulos2023new}. In recent years, modified gravity theories have garnered significant attention as potential solutions to outstanding problems in cosmology and astrophysics, such as the nature of dark energy and dark matter, and the late-time accelerated expansion of the Universe. Among these theories, $f(Q,T)$ gravity presents a novel approach by incorporating both the non-metricity scalar $Q$ and the trace of the energy-momentum tensor $T$. This framework extends the scope of modified gravity theories by introducing new interactions between matter and geometry. $f(Q,T)$ gravity modifies the geometric structure of spacetime by considering a function $f$ of both $Q$, which characterizes the non-metricity of spacetime, and $T$, representing the matter content. This theory stands out because it generalizes the metric-affine theories and provides a rich structure that can encapsulate various gravitational phenomena under a single framework. By incorporating the trace of the energy-momentum tensor, $f(Q,T)$ gravity can potentially offer insights into how matter influences the geometry of spacetime in ways not accounted for in General Relativity (GR) or other modified gravity theories. Yixin et al. \cite{25} examined the cosmological implications of the $f(Q,T)$ theory across three specific model classes. The resulting solutions account for both accelerating and decelerating phases of the Universe's evolution, suggesting that $f(Q,T)$ gravity offers valuable insights into both early and late stages of cosmological evolution. Despite being relatively new, the $f(Q,T)$ gravity theory has found intriguing applications in the literature. Arora et al. \cite{26:2020tuk,27,28} have investigated the viability of the $f(Q, T)$ gravity as a method to explain the present and late-time cosmic acceleration and provide a feasible solution to the dark energy problem. They also tested the $f(Q,T)$ gravity theory using energy conditions \cite{Arora_2021E,Arora_2020E}, and the results enabled them to verify the viability of various families of $f(Q,T)$ gravity models. This opens up new possibilities for a comprehensive description of gravity that is compatible with the dark energy era and incorporates effects from the quantum era of the Universe. In the presence of $f (Q, T)$ theory, gravitational baryogenesis has been studied in \cite{29}. Also, the cosmological perturbation theory in the framework of $f(Q, T)$ theory has been studied in \cite{30}. Gadbail et al. \cite{31} have reconstructed the $f(Q, T)$ models for various cosmological scenarios and show the model's validity. In Ref. \cite{Loo_2023}, the authors present a fully covariant formulation of the $f(Q,T)$ theory. Utilizing this formulation, they construct the correct energy balance equation and perform a dynamical system analysis within a spatially flat FLRW spacetime. Additionally, in $f(Q,T)$ gravity, some work is done on the astrophysical object \cite{32,33}. These studies make $f(Q,T)$ gravity a compelling candidate for further theoretical and observational investigation.\\\\
In this paper, we look at a specific type of $f(Q,T)$ theory, in which the non-metricity $Q_{\sigma\mu\nu}$ of the space-time is depicted in its standard Weyl form, and it is completely determined by a vector field $w_{\mu}$. More precisely, we make use of Weyl geometry and explicitly express non-metricity as $Q_{\sigma\mu\nu}=2\,w_{\sigma}\,g_{\mu\nu}$.
We will also look at the flat geometry restraint, which requires that the Weyl curvature scalar vanish. This condition is added to the action through a  Lagrange multiplier $\lambda$. 
Furthermore, the field equation of Weyl-type $f(Q, T)$ gravity is obtained by varying the action with respect to metric tensor together with a vector field. Yixin et al. \cite{34} studied the cosmological implication of Weyl-type $f(Q,T)$ gravity for three types of $f(Q,T)$ model and obtained solutions showing both the decelerating and accelerating phase of the Universe. They demonstrated that the Weyl-type $f(Q,T)$ theory is an alternate and useful strategy for describing the early and late stages of cosmological evolution. Yang et al. \cite{35} studied the  Raychaudhuri equation, Geodesic deviation, tidal force, and Newtonian limit in Weyl-type $f(Q, T)$ theory. Furthermore, the power law and viscous cosmology were studied in Weyl-type $f(Q, T)$ theory \cite{36,37}. Weyl-type $f(Q, T)$ theory has lately undergone some more work, which is described in references \cite{gadbail2022interaction,gadbail2023dark}.
The main goal of the present study is to examine the Weyl-type $f(Q, T)$ theory by using the barotropic equation of state $p=w\rho$. For this study, we took two type of $f(Q,T)$ models: (I) $f(Q,T)=\alpha Q+\frac{\beta}{6\kappa^2}T$ (linear) and (II) $f(Q,T)=\alpha Q^{m+1}+\frac{\beta}{6\kappa^2}T$ (nonlinear). After that, we find two solutions to the field equation for both linear and nonlinear models. We further analyze the obtained solutions statistically using observational datasets. We observe that some cosmological parameters, like equation-of-state and deceleration parameters, are crucial in defining the cosmic evolution of the Universe. And it is well comprehended that the equation-of-state parameter implies a variety of fluid Universe descriptions. As a result, using observational data to restrict this parameter is intriguing. For this purpose, we use the $Pantheon^+$ dataset, which contains 1701 data points. Later on,  we statistically compare our models to the $\Lambda$CDM model using ${{\protect\chi}^2_{min}}$, ${{\protect\chi}^2_{red}}$, $AIC$, $\Delta AIC$, $BIC$  and $\Delta BIC$. In order to identify the dark energy profile of our model, we also test it using the Om diagnostic and compare it to the $\Lambda$CDM model.\\\\ 
The novelty of this work lies in the detailed exploration of Weyl-type $f(Q, T)$ gravity with a barotropic equation of state, focusing on both linear and nonlinear models. By statistically analyzing the resulting solutions using the extensive Pantheon+ dataset and comparing them to the $\Lambda$CDM model, we aim to provide new insights into the dark energy profile and cosmic evolution. This study will help to further validate Weyl-type $f(Q, T)$ theories as viable alternatives for describing the late-time acceleration expansion of the Universe.\\\\
The layout of the article is as follows. In section \ref{section 2}, we present the basic formulation of Weyl-type $f(Q,T)$ gravity. In section \ref{section 3}, we find the solution of the generalized Friedmann equation by using the barotropic EoS parameter $p=w\rho$ for both linear and nonlinear models. In section \ref{section 4}, we study the statistical analysis of Weyl $f(Q,T)$ models by using the $Pantheon^+$ dataset and present our results. In section \ref{section 5}, we test our model using $Om$ diagnostic and compare it to the $\Lambda$CDM model to determine its dark energy profile. In section \ref{section 6}, we conclude our final results.
\section{Basic formulation of Weyl-type $f(Q,T)$ theory}\label{section 2}
In 1918, Weyl proposed an essential generalization of Riemannian geometry, which served as the mathematical foundation for general relativity, by assuming that during the parallel transport of vectors, both the orientation and magnitude of a vector change \cite{38}. Weyl introduced the intrinsic vector field $\omega_{\mu}$ and a semi-metric connection to express the simultaneous change of direction and length,


\begin{equation}
\label{1}
\widetilde{\Gamma}^\lambda_{\,\,\,\mu\nu}\equiv\Gamma^\lambda_{\,\,\,\mu\nu}+g_{\mu\nu}\omega^\lambda-\delta^\lambda_\mu \omega_\nu-\delta^\lambda_\nu \omega_\mu,
\end{equation}
where tilde shows the quantities defined in Weyl geometry, and the Christoffel symbol $\Gamma^\lambda_{\,\,\,\mu\nu}=\frac{1}{2}g^{\lambda\sigma}\left(\partial_{\mu}g_{\sigma\nu}+\partial_{\nu}g_{\sigma\mu}-\partial_{\sigma}g_{\mu\nu}\right)$.\\
 
With the help of a semi-metric connection, one can obtain the result in Weyl geometry,
\begin{equation}
\label{2}
\widetilde{\nabla}_{\sigma}g_{\mu \nu}= 2\omega_{\sigma}g_{\mu\nu}.
\end{equation}
The scalar non-metricity is defined as  

\begin{equation}
\label{3}
Q\equiv- g^{\mu\nu}\left(L^\alpha_{\,\,\,\beta\nu}L^\beta_{\,\,\,\nu\alpha}-L^\alpha_{\,\,\,\beta\alpha}L^\beta_{\,\,\,\mu\nu}\right),
\end{equation}
where $L^\lambda_{\,\,\,\mu\nu}$ is the disformation tensor read as
\begin{equation}
\label{4}
L^\lambda_{\mu\nu}=-\frac{1}{2}g^{\lambda\gamma}\left(Q_{\mu\gamma\nu}+Q_{\nu\gamma\mu}-Q_{\gamma\mu\nu}\right).
\end{equation}
The nonmetricity tensor $Q_{\lambda\mu\nu}$ is defined as the covariant derivative of the metric tensor with respect to the semi-metric connection $\widetilde{\Gamma}_{\,\,\,\mu\nu}^{\lambda}$,
\begin{equation}
\label{5}
Q_{\lambda\mu\nu}\equiv\widetilde{\nabla}_\lambda g_{\mu\nu}=\partial_\lambda g_{\mu\nu}-\widetilde{\Gamma}^\rho_{\,\,\,\lambda\mu}\,g_{\rho\nu}-\widetilde{\Gamma}^\rho_{\,\,\,\lambda\nu}\,g_{\rho\mu}=2\omega_\lambda g_{\mu\nu}.
\end{equation}\\
Inserting  Eq. \eqref{5} in Eqs. \eqref{3} and \eqref{4}, we obtained the important relation 
\begin{equation}
\label{6}
Q=-6\omega^2.
\end{equation}

 The action of Weyl-type $f(Q,T)$ theory of gravity coupled with matter Lagrangian $\mathcal{L}_m$ is provided by \cite{34}
 \begin{multline}
\label{7}
 S=\int d^4x\sqrt{-g}\left[ \kappa^2f(Q,T)-\frac{1}{2}\Tilde{m}^2\omega_\mu \omega^\mu-\frac{1}{4}W_{\mu\nu}W^{\mu\nu}+\right.\\ 
 \left. \lambda\left(R+6\nabla_\alpha \omega^\alpha-6\omega_\alpha \omega^\alpha\right)+\mathcal{L}_m\right],
\end{multline}    
with  $\kappa^2=\frac{1}{16\pi G}$. Here, $f(Q, T)$ is an arbitrary function of $Q$ (non-metricity scalar) and $T$ (energy-momentum tensor trace). The particle's mass to the vector field $\omega_{\mu}$ is denoted by $\Tilde{m}$. The second and third terms in action are the mass term and ordinary kinetic term of $\omega_{\mu}$, respectively. Furthermore, under the assumption of total scale curvature vanishing, a condition is introduced to the action through a Lagrange multiplier $\lambda$.  \\
Further, the generalized Proca equation by varying the action \eqref{7} with respect to a vector field is 
\begin{equation}
\label{8}
\nabla^\nu W_{\mu\nu}-\left(\Tilde{m}^2+12\kappa^2f_Q+12\lambda\right)\omega_\mu=6\nabla_\mu \lambda.
\end{equation}
We obtained the generalized field equation by varying the action \eqref{7} with regard to the metric tensor together with a vector field,

\begin{multline}
\label{9}
\frac{1}{2}\left(T_{\mu\nu}+S_{\mu\nu}\right)-\kappa^2f_T\left(T_{\mu\nu}+\Theta_{\mu\nu}\right)=-\frac{\kappa^2}{2}g_{\mu\nu}f\\-6k^2f_Q \omega_\mu \omega_\nu +
\lambda\left(R_{\mu\nu}-6\omega_\mu \omega_\nu +3g_{\mu\nu}\nabla_\rho \omega^\rho \right)
\\+3g_{\mu\nu}\omega^\rho \nabla_\rho \lambda 
-6\omega_{(\mu}\nabla_{\nu )}\lambda+
g_{\mu\nu}\square \lambda-\nabla_\mu\nabla_\nu \lambda,
\end{multline}
in which 
 \begin{equation}
\label{10}
f_Q\equiv\frac{\partial f(Q,T)}{\partial Q},
 \hspace{0.2in}
f_T\equiv \frac{\partial f(Q,T)}{\partial T}.
\end{equation}
Also, the definition of $T_{\mu\nu}$ and $\Theta_{\mu\nu}$ as
\begin{equation}
\label{11}
T_{\mu\nu}\equiv-\frac{2}{\sqrt{-g}}\frac{\delta(\sqrt{-g}L_m)}{\delta g^{\mu\nu}},
\end{equation} 

\begin{equation}
\label{12}
\Theta_{\mu\nu}=g^{\alpha\beta}\frac{\delta T_{\alpha\beta}}{\delta g_{\mu\nu}}=g_{\mu\nu}L_m-2T_{\mu\nu}-2g^{\alpha\beta}\frac{\delta^2 L_m}{\delta g^{\mu\nu}\delta g^{\alpha\beta}}.
\end{equation}
Here, the re-scaled energy-momentum tensor of the free Proca field is denoted by $S_{\mu\nu}$ and is defined as
\begin{equation}
\label{13}
S_{\mu\nu}=-\frac{1}{4}g_{\mu\nu}W_{\rho\sigma}W^{\rho\sigma}+W_{\mu\rho}W^\rho_\nu -\frac{1}{2}\Tilde{m}^2g_{\mu\nu}\omega_\rho \omega^\rho +\Tilde{m}^2 \omega_\mu \omega_\nu ,
\end{equation}
and 
\begin{equation}
\label{14}
W_{\mu\nu}=\nabla_\nu \omega_\mu-\nabla_\mu \omega_\nu.
\end{equation}
It is also noted that the expression for the divergence of the matter energy-momentum tensor in the Weyl-type $f(Q,T)$ theory is given by \cite{34}
\begin{equation*}
\nabla^{\mu}T_{\mu\nu}= \frac{\kappa^2}{1+2\kappa^2f_T}\left[2\nabla_{\nu}(f_T\mathcal{L}_m)-f_T\nabla_{\nu}T-2T_{\mu\nu}\nabla^{\mu}f_T\right].  
\end{equation*}
\section{Solution of the FLRW model} 
\label{section 3}
The standard Friedmann-Lemaitre-Robertson-Walker line element is taken into account in the present study and is written explicitly as
\begin{equation}
\label{15}
ds^2=-dt^2+a^2(t)\delta_{ij}dx^i dx^j ,
\end{equation}
where $a(t)$ is the cosmic scale factor.\\
The vector field  $\omega_\mu$ is assumed to have the following form in the spatial symmetry: $\omega_\mu=\left[\psi (t),0,0,0\right]$ \cite{34} implying $\omega^2=\omega_\mu \omega^\mu=-\psi^2(t)$, and  $Q=-6\omega^2=6\psi^2(t)$.\\\\
The energy-momentum tensor for the perfect fluid is defined as:
\begin{equation}
\label{16}
T_{\mu\nu}=\left(\rho+p\right)u_\mu u_\nu+ p g_{\mu\nu},
\end{equation}
where $p$ and $\rho$ are the pressure and the matter-energy density, respectively. The four-velocity vector $u^\mu$ is such that $u_\mu u^\mu=-1$. Thus implies  $T^\mu_\nu=diag\left(-\rho,p,p,p\right)$, and $\Theta^\mu_\nu=\delta^\mu_\nu p-2T^\mu_\nu=diag\left(2\rho+p,-p,-p,-p\right)$. 
The flat space constraint and the generalized Proca equation in the cosmological case can be represented as
\begin{equation}
\label{17}
\dot{\psi}=\dot{H}+\psi^2+2H^2-3H\psi,
\end{equation}
\begin{equation}
\label{18}
\dot{\lambda}=\left(-\frac{1}{6}\Tilde{m}^2-2\kappa^2f_Q-2\lambda\right)\psi=-\frac{1}{6}\Tilde{m}^2_{eff}\psi ,
\end{equation}
\begin{equation}
\label{19}
\partial_i \lambda=0.
\end{equation}
From Eq \eqref{9} and using given metric \eqref{15} the obtained modified Friedmann equation is,
\begin{multline}
\label{20}
\frac{1}{2}\rho+\kappa^2f_T\left(p+\rho\right)=\frac{\kappa^2}{2}f-\left(6\kappa^2f_Q\,\psi^2+\frac{1}{4}\Tilde{m}^2\psi^2\right) \\
+3\lambda\left(H^2-\psi^2\right)+3\dot{\lambda}\left(H-\psi\right),
\end{multline}

\begin{multline}
\label{21}
-\frac{1}{2}p=\frac{\kappa^2}{2}f+\frac{\Tilde{m}^2\psi^2}{4}+\lambda\left(3H^2+3\psi^2+2\dot{H}\right)\\
+\left(2H+3\psi\right)\dot{\lambda}+\ddot{\lambda}.
\end{multline}
where $\rho$ and $p$ are the matter-energy density and the pressure, respectively. dot (.) represents the derivative with respect to time $t$. \\
By using the above Friedmann equations, the energy balance equation can be obtained as
\begin{equation}
    \dot{\rho}+3H(\rho+p)=\frac{1}{1+2\kappa^2\,f_T}\left[2\kappa^2(\rho+p)\dot{f_T}-f_T(\dot{\rho}-\dot{p})\right].
\end{equation}
As a result, the above equation shows that the matter
energy-momentum tensor is not conserved in the Weyl-type $f(Q, T)$ theory. The non-conservation of the matter energy-momentum tensor can be interpreted physically as indicating the presence of an extra force acting on massive test particles, causing the motion to be nongeodesic. From a physical perspective, it indicates the amount of energy that enters or leaves a specified volume of a physical system. Moreover, the non-vanishing right-hand side of the energy-momentum tensor indicates the transfer processes or particle production in the system. One can note that the energy-momentum tensor becomes conserved in the absence of $f_T$ terms in the above equation.\\
We study the above Friedmann equation \eqref{20} and \eqref{21} for two type of $f(Q,T)$ models: (I) $f(Q,T)=\alpha Q+\frac{\beta}{6\kappa^2}T$ (linear) and (II) $f(Q,T)=\alpha Q^{m+1}+\frac{\beta}{6\kappa^2}T$ (nonlinear).
\subsection{Model I :- $f(Q,T)=\alpha Q+\frac{\beta}{6\kappa^2}T$ }
For our investigation, we consider the linear functional form $f(Q,T)=\alpha Q+\frac{\beta}{6\kappa^2}T$, where $\alpha$ and $\beta$ are model parameters. This particular functional form of $f(Q,T)$ is motivated in reference \cite{34}. For certain choices of model parameters, this model is basically equivalent to $\Lambda$CDM model for certain redshift ranges. The choice of these models stems from their relevance within the context of addressing late-time phenomenology. Our primary focus was on investigating alternatives to the standard cosmological framework. The power law model, as derived from our first condition, offers a promising candidate for addressing the inflation issue on further investigation. Using this form and $\lambda=\kappa^2$ (for special case), we rewrite the field equations \eqref{20} and \eqref{21} as
\begin{equation}
\label{22}
-\left(\frac{\beta}{4}+\frac{1}{2}\right)\rho+\frac{\beta}{12}p=3\alpha \psi^2+\frac{\mathcal{M}^2\psi^2}{4}+3\left(\psi^2-H^2\right),
\end{equation}
\begin{multline}
\label{23}
-\left(\frac{\beta}{4}+\frac{1}{2}\right)p+\frac{\beta}{12}\rho=3\alpha \psi^2+\frac{\mathcal{M}^2\psi^2}{4} \\ +\left(3\psi^2+3H^2+2\dot{H}\right),
\end{multline}
where $\mathcal{M}^2= \Tilde{m}^2/\kappa^2$, $M$ represents the mass of the Weyl vector field and indicates the intensity of the Weyl geometry-matter coupling. In this situation, we've assumed that $\mathcal{M}= 0.90$ \cite{34}. To proceed further, we consider the relation of pressure and energy density as $p=w \rho$. Using the relation $\nabla_{\lambda}g_{\mu\nu}=-\omega_{\lambda}g_{\mu\nu}$ and $\omega_1=\psi(t)$, we obtained $\psi(t)=-6H(t)$. Further simplifying Eq.\eqref{22} and \eqref{23}, we obtained the first order differential equation as 
\begin{equation}
    A\,H(z)-B\,(1+z)\frac{dH(z)}{dz}=0,
\end{equation}
where $A=\frac{36\alpha+3\mathcal{M}^2+35}{3\beta-w\beta+6}-\frac{36\alpha+3\mathcal{M}^2+37}{6w-\beta+3w\beta}$ and $B=\frac{2}{9(2+\beta)w-3\beta}$.\\
The solution of the above differential equation is 
\begin{equation}
\label{25}
H(z)=H_0 (1+z)^{\mathcal{X}},
\end{equation}
where
\begin{multline*}
    \mathcal{X}=\frac{3w(105+71\beta+36\alpha(3+2\beta))}{\beta  (w-3)-6}\\
    +\frac{3\left(-111-108\alpha-73\beta-72\alpha\beta+3\mathcal{M}^2(-1+w)(3+2\beta)\right)}{\beta  (w-3)-6}.
\end{multline*}
The deceleration parameter $q$ is another important cosmological quantity and it is defined as (in terms of redshift $z$)
\begin{equation}
\label{26}
    q(z)=-1+(1+z)\frac{1}{H(z)}\frac{dH(z)}{dz}.
\end{equation}
After inserting Eq. \eqref{25} in \eqref{26}, we obtained the deceleration parameter for our model I as 
\begin{multline}
    \label{27}
q(z)=-1+\frac{3\left(w (36 \alpha  (2 \beta +3)+71 \beta +105)-111\right)}{\beta  (w-3)-6}\\
+\frac{3 \left(-72 \alpha  \beta -108 \alpha -73 \beta +3 (2 \beta +3) \mathcal{M}^2 (w-1)\right)}{\beta  (w-3)-6}.
\end{multline}
In the linear case, we found a power-law solution to the Friedmann equations. A fascinating answer to several unusual problems, including flatness and the horizon problem, is power-law cosmology. The literature has a strong justification for the power-law cosmology. Kumar \cite{39} examined cosmic parameters using power law and data from Hz and SNe Ia. Statefinder analysis was used by Rani et al. \cite{40} to analyze the power-law cosmology. In power-law solution scenario, the expression of the deceleration parameter is constant. However, due to the constant value of the deceleration parameter, it fails to offer redshift transition from deceleration to acceleration. 
\subsection{Model II :- $f(Q,T)=\alpha Q^{m+1}+\frac{\beta}{6\kappa^2}T$}
For our investigation, we also consider the non-linear functional form $f(Q,T)=\alpha Q^{m+1}+\frac{\beta}{6\kappa^2}T$, where $\alpha$, $\beta$ and $m$ are free model parameters. It is worth mentioning that $m=0$, $\beta=0$, and $\alpha=1$ correspond to a case of the successful theory of general relativity (GR). But, when $m\neq 0$ and $\beta=0$ it corresponds to the $f(Q)$ gravity, which is not equivalent to GR. To study the $f(Q, T)$ model's non-linear case, we try to constrain $m\neq 0$ through observational data, which is not equivalent to GR. Arora et al. \cite{26:2020tuk} examined late time cosmology using this sort of model by limiting the free parameters with observational data sets of the updated 57 points of Hubble data sets and 580 points of union 2.1 compilation supernovae data sets. Consideration of inflation in the presence of scalar fields is another possible application of this type of model, and it may offer a completely different viewpoint on the geometrical, gravitational, and cosmological processes that did play a significant role in the very early dynamics of the Universe.\\
Using this non-linear functional form and $\lambda=\kappa^2$ (for special case), we rewrite the field equations \eqref{20} and \eqref{21} as
\begin{multline}
\label{28}
-\left(\frac{\beta}{4}+\frac{1}{2}\right)\rho+\frac{\beta}{12}p=\left(6m+3\right)\alpha\,6^m\,\psi^{2m+2}\\
+\frac{1}{4}\mathcal{M}^2\psi^2+3(\psi^2-H^2),
\end{multline}

\begin{multline}
\label{29}
-\left(\frac{\beta}{4}+\frac{1}{2}\right)p+\frac{\beta}{12}\rho=3\,\alpha\, 6^m\, \psi^{2m+2}+\frac{\mathcal{M}^2\psi^2}{4}\\
+\left(3\psi^2+3H^2+2\dot{H}\right).
\end{multline}
To proceed further, we consider the relation of pressure and energy density as $p=w \rho$. Using the relation $\nabla_{\lambda}g_{\mu\nu}=-\omega_{\lambda}g_{\mu\nu}$ and $\omega_1=\psi(t)$, we obtained $\psi(t)=-6H(t)$.
Further simplifying Eq.\eqref{28} and \eqref{29}, we obtained the first order differential equation as 
\begin{equation}
\label{30}
    A\,H^{2m+1}(z)+B\,H(z)+C\,(1+z)\,\frac{dH(z)}{dz}=0,
\end{equation}
where $A=\frac{36 \alpha  (2 m+1) 6^{3 m+2}}{-3 \beta +\beta  w-6}-\frac{36 \alpha  6^{3 m+2}}{\beta -3 \beta  w-6 w}$, $B=\frac{108 \mathcal{M}^2+1260}{-3 \beta +\beta  w-6}-\frac{108 \mathcal{M}^2+1332}{\beta -3 \beta  w-6 w}$, and $C=\frac{24}{\beta -3 \beta  w-6 w}$ are constants.

The solution of the above differential equation is 
\begin{equation}
\label{31}
H(z)=-\frac{A}{B}+\left(H_0+\frac{A}{B}\right)\left(1+z\right)^{\frac{2\,m\,B}{C}}.
\end{equation}
After inserting Eq. \eqref{31} in \eqref{26}, we obtained the deceleration parameter for model II as
\begin{equation}
    q(z)=-1+\frac{2\,m\,B \left(\frac{A}{B}+H_0\right) (z+1)^{\frac{2\,m\,B}{C}}}{C \left(\left(\frac{A}{B}+H_0\right) (z+1)^{\frac{2\,m\,B}{C}}-\frac{A}{B}\right)}.
\end{equation}
In the nonlinear case, we found a hybrid solution to the Friedmann equations. In this scenario, the expression of the deceleration parameter is non-constant. As a result, for this instance, we can offer a redshift transition from deceleration to acceleration. 

\section{Observational data, MCMC results, and information criteria}
\label{section 4}
The most favorable Weyl $f(Q,T)$ models are taken into consideration in this work, and they are evaluated using various combinations of observational data sets. We use the publically accessible $emcee$ software, which is available at Ref. \cite{41}, to conduct an MCMC (Monte Carlo Markov Chain) analysis for each Weyl $f(Q,T)$ model and data set combination. By altering the parameters in a variety of conservative priors and examining the posteriors of the parameter space, the MCMC sampler constrains the model and cosmological parameters. As a result, we obtain the one- and two-dimensional distributions for each parameter, with the one-dimensional distribution representing the posterior distribution of the parameter and the two-dimensional distribution showing the covariance between two different values. In the present section, we analyze  observational datasets that will be used in our research, as well as the statistical methods that will be applied. We will specifically employ data from Type Ia supernovae. Further, we propose several information criteria that provide information on the fit's quality.\\\\
\subsection*{$Pantheon^+$ Sample (Without SHOES Calibrated)}
One of the most recent compilations of spectroscopically confirmed Type Ia supernovae (SNeIa), which we call $Pantheon^+$; this sample is a direct inheritor of the Pantheon analysis \cite{42}, which itself succeeded the Joint Light-curve Analysis.  In terms of cosmology, the fundamental distinction between the $Pantheon^+$ study and the original Pantheon analysis is the inclusion of fresh data sets in the latter. The $Pantheon^+$ analysis includes 1701 SNIa samples, up from 1048 in the original Pantheon analysis. The 1701 light curves of 1550 unique Type Ia supernovae (SNe Ia) in the redshift range $0.001 < z < 2.26$ make up the $Pantheon^+$ dataset \cite{43,44}. The constraining power of SNeIa evolves manifest when employed as standard candles. This can be accomplished by using the distance modulus:
\begin{equation}
\mu_{th}= 5log\left( \frac{d_{L}(\textbf{x},z)}{Mpc}\right) + 25 + M , 
\end{equation}
where $d_{L}$ is the luminosity distance and it can be written as 
\begin{equation}
d_{L}(\textbf{x},z)= c(1+z)\int_{0}^{z} \frac{dz'}{H(\textbf{x},z')}.
\end{equation}
For the Pantheon sample, the chi-squared function $\chi^2_{SN}$ is defined as
\begin{equation}
\chi^{2}_{SN}= \mu_{SN} C^{-1} \mu^{T}_{SN},
\end{equation}
where $\mu_{SN}= \mu_{i}-\mu_{th}(\theta_{s},z_{i})$ and $\mu_{i}=\mu_{B,i}-M$. The apparent maximum magnitude for redshift $z_i$ is represented here by $\mu_{B,i}$. The hyper-parameter $M$ measures uncertainties from many sources. It is utilized instead of free parameters space $(H_0,\alpha,\beta,w,M)$ and $(H_0,\alpha,\beta,m,w,M)$  in the view of the "BEAMS" with Bias Corrections method \cite{48}. 
\begin{table*}
\centering
\begin{tabular}{|c|c|c|c|c|c|}
\hline
\multicolumn{4}{|c|}{MCMC Results} \\
\hline\hline
Model & Parameters & Priors & Best fit Value \\[1ex]
\hline
$  \Lambda$CDM Model  & $H_0$ & [50,100] &$74.035632_{\pm 1.402626}^{\pm 2.835751}$ \\[1ex]
&$\Omega_{m0}$ & [0.,1.] &$0.285121_{\pm 0.007998}^{\pm 0.015325}$ \\[1ex]
&M & [-19.,-20.0] &$-19.233938_{\pm 0.041619}^{\pm 0.083213
}$ \\[1ex]
\hline
Model I & $H_0$& [50,100] & $73.545070_{\pm 1.512298}^{\pm 2.539369}$ \\[1ex] 
&$\alpha$&[-2.,-1.]  & $-1.057789_{\pm 0.001626}^{\pm 0.002503}$ \\[1ex]
&$\beta$&[0.,1.] & $0.464107_{\pm 0.390927}^{\pm 0.605504}$  \\[1ex]
&$w$ & [-2.,-1.] & $-1.480871_{\pm 0.222890}^{\pm 0.310929}$ \\[1ex]
& $M$ & [-19.,-20.0] & $-19.221769_{\pm 0.043751}^{\pm 0.077376}$ \\[1ex]
\hline
Model II & \(H_0\) & [50, 100] & \(74.134762_{\pm 1.097884}^{\pm 1.643197}\) \\[1ex]
& \(\alpha\) & [-1., 0.] & \(-0.577999_{\pm 0.186027}^{\pm 0.272576}\) \\[1ex]
& \(\beta\)  & [-2., -1.] & \(-1.533206_{\pm 0.005372}^{\pm 0.008119}\) \\[1ex]
& \(m\) & [0., 1.]  & \(0.388177_{\pm 0.034718}^{\pm 0.048863}\) \\[1ex]
& \(w\) & [-0.1, 0.5] & \(-0.114274_{\pm 0.078364}^{\pm 0.154546}\) \\[1ex]
& \(M\) & [-19.0, -20.0] & \(-19.238416_{\pm 0.033487}^{\pm 0.051279}\) \\[1ex]
\hline
\end{tabular}
\caption{Summary of the mean parameter values and their uncertainties at the 68\% confidence and 95\% confidence level for the $\Lambda$CDM, Model I, and Model II cosmological models}\label{tab_MCMC}
\label{table_1}
\end{table*}
\begin{figure*}
\centering
\includegraphics[scale=1.2]{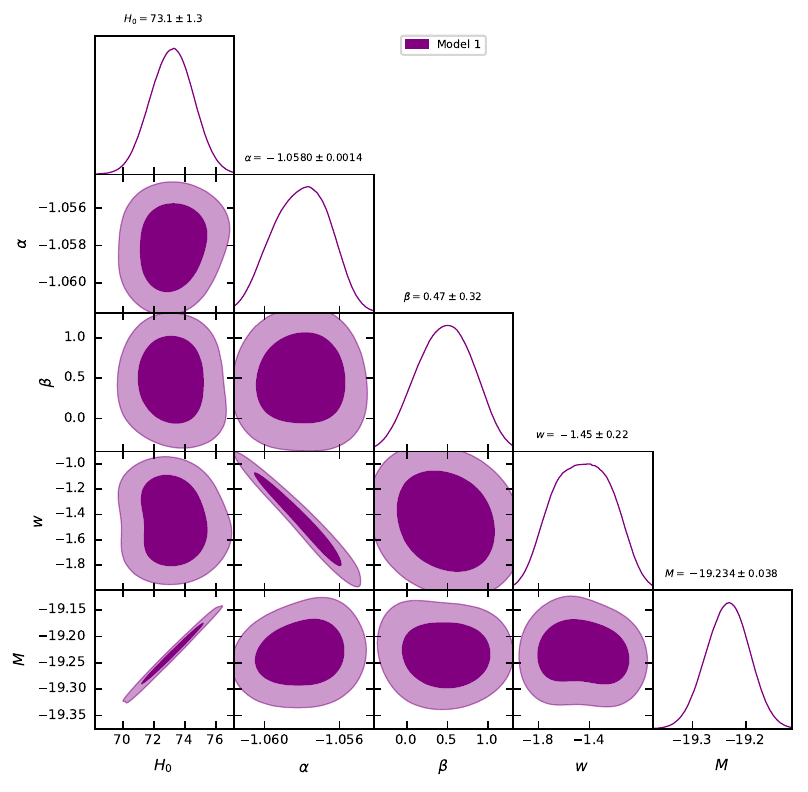}
\caption{The contour plot of free parameter space $(H_0,\alpha,\beta,w,M)$ for our model I with $1\sigma$ and $2\sigma$ errors obtained from the  datasets.}\label{fig_1}
\end{figure*}
\begin{figure*}
\centering
\includegraphics[scale=1.2]{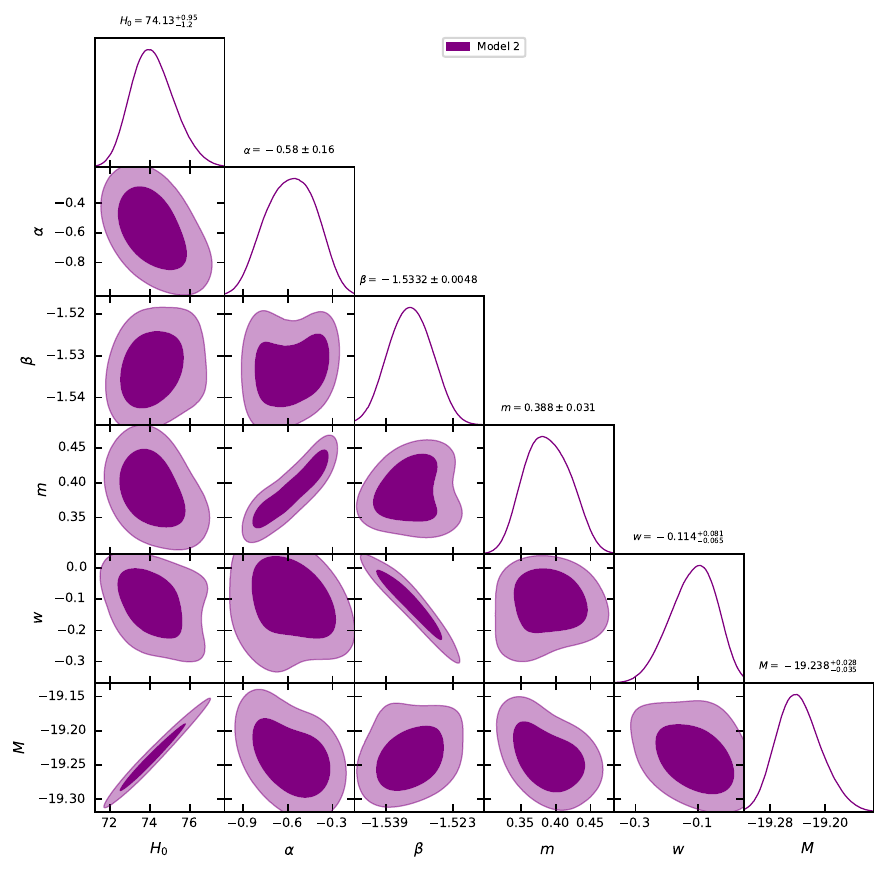}
\caption{The contour plot of free parameter space $(H_0,\alpha,\beta,m,w,M)$ for our model II with $1\sigma$ and $2\sigma$ errors obtained from the  datasets.}\label{fig_2}
\end{figure*}
\begin{figure*}
\begin{subfigure}{.48\textwidth}
\includegraphics[width=\linewidth]{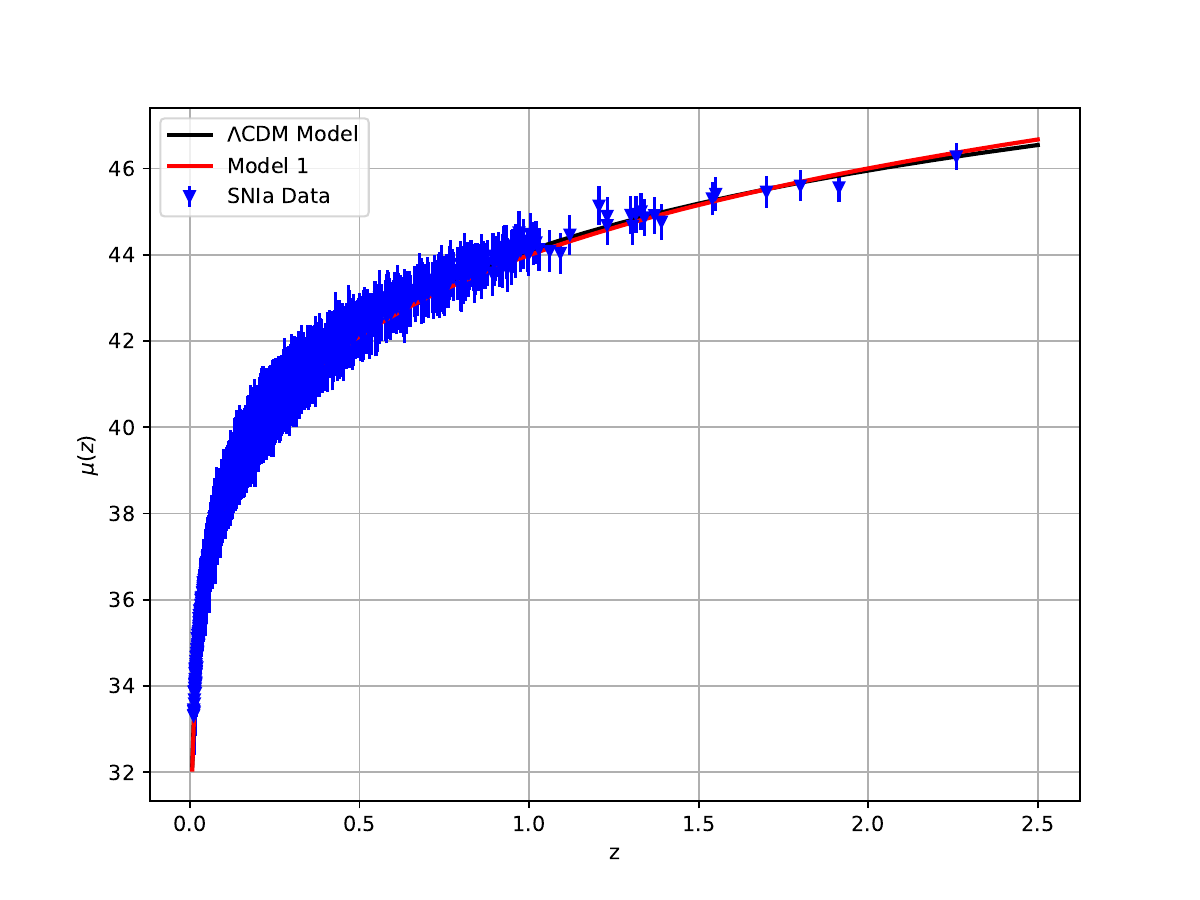}
    \caption{Plot of $\mu$ of Model I.}
    \label{fig_3a}
\end{subfigure}
\hfil
\begin{subfigure}{.48\textwidth}
\includegraphics[width=\linewidth]{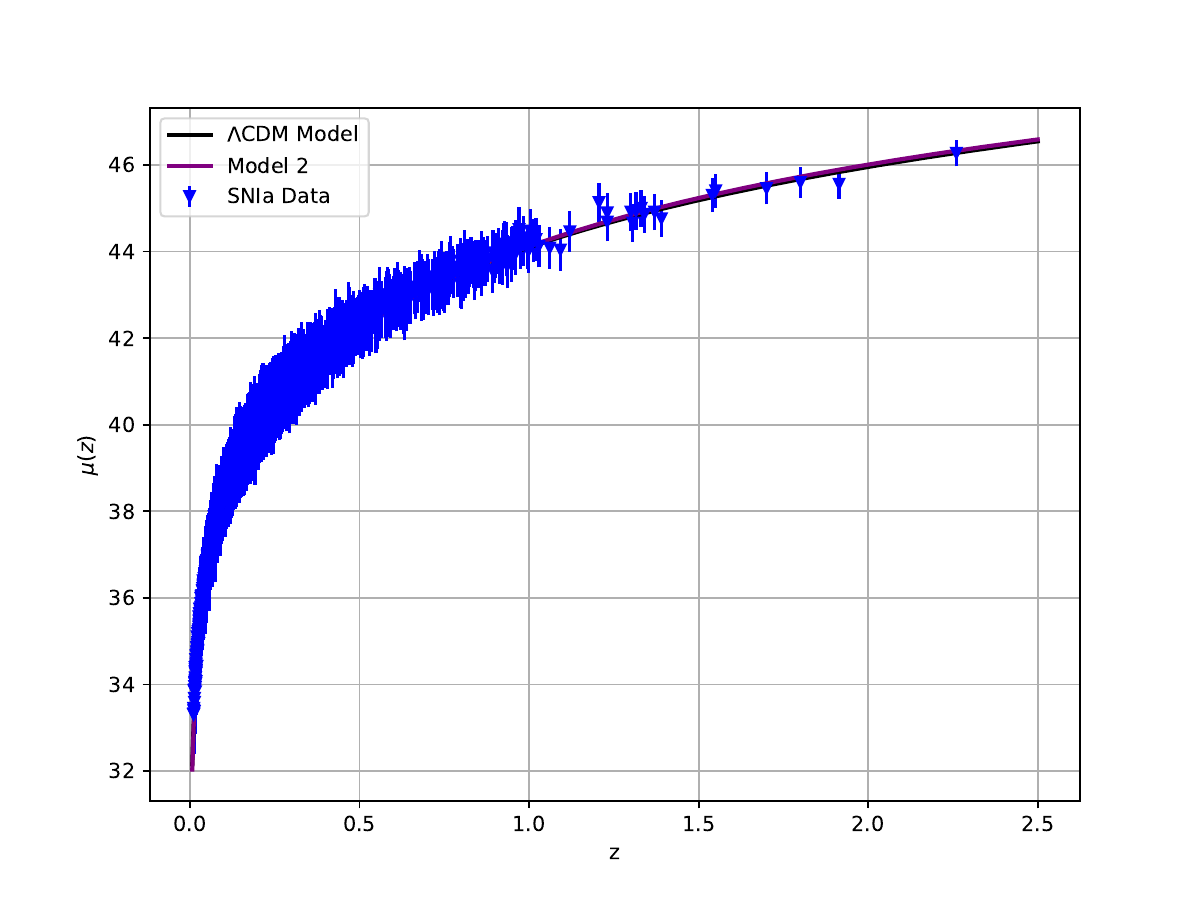}
    \caption{Plot of $\mu$ of Model II.}
    \label{fig_3b}
\end{subfigure}\hfil
\caption{The figure shows the theoretical curve of the distance modulus \( \mu(z) \) for \(\Lambda\)CDM (black line), along with Model 1 (red line) and Model 2 (purple line), compared against the Supernova Type Ia dataset (blue points with error bars). For the \(\Lambda\)CDM model, the parameters were fixed at \(\Omega_{\mathrm{m0}} = 0.3\) and \(\Omega_\Lambda = 0.7\).}\label{fig_3}
\end{figure*}
\begin{figure*}
\begin{subfigure}{.48\textwidth}
\includegraphics[width=\linewidth]{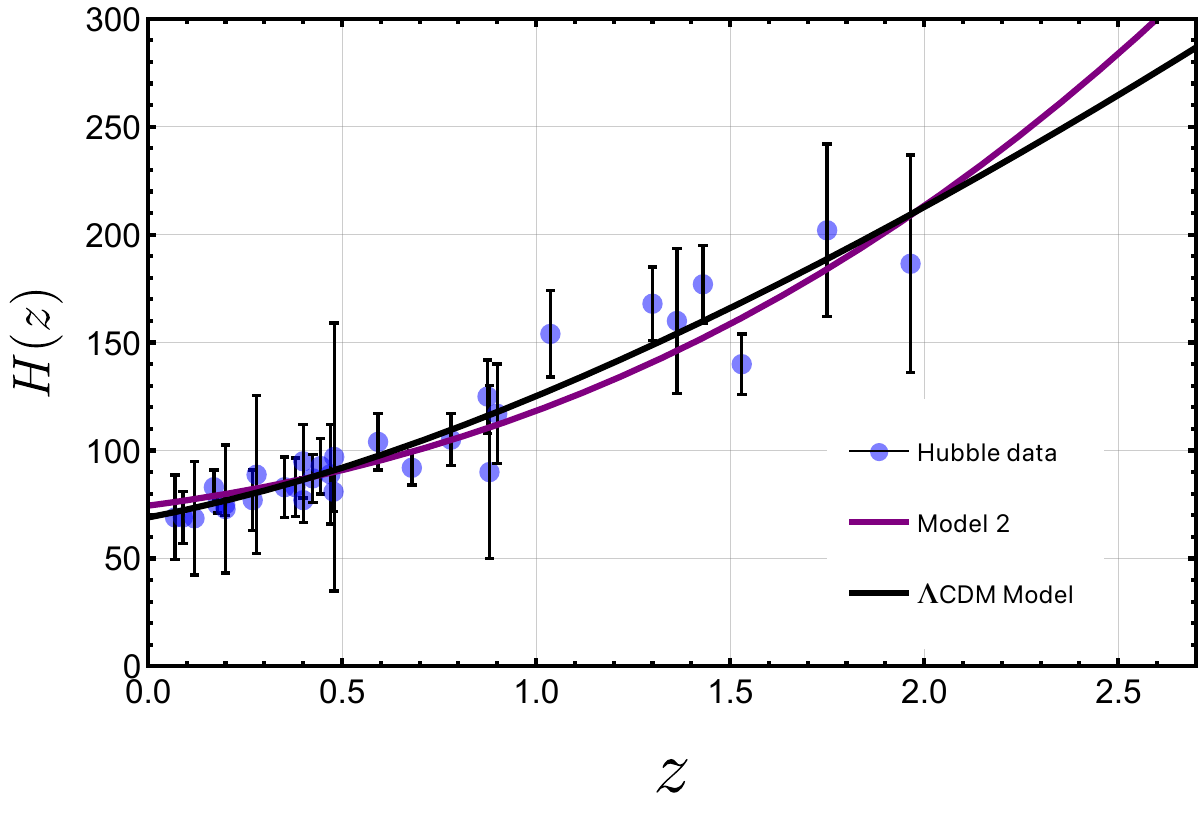}
    \caption{Plot of $H(z)$ parameter of Model I.}
    \label{fig_3a}
\end{subfigure}
\hfil
\begin{subfigure}{.48\textwidth}
\includegraphics[width=\linewidth]{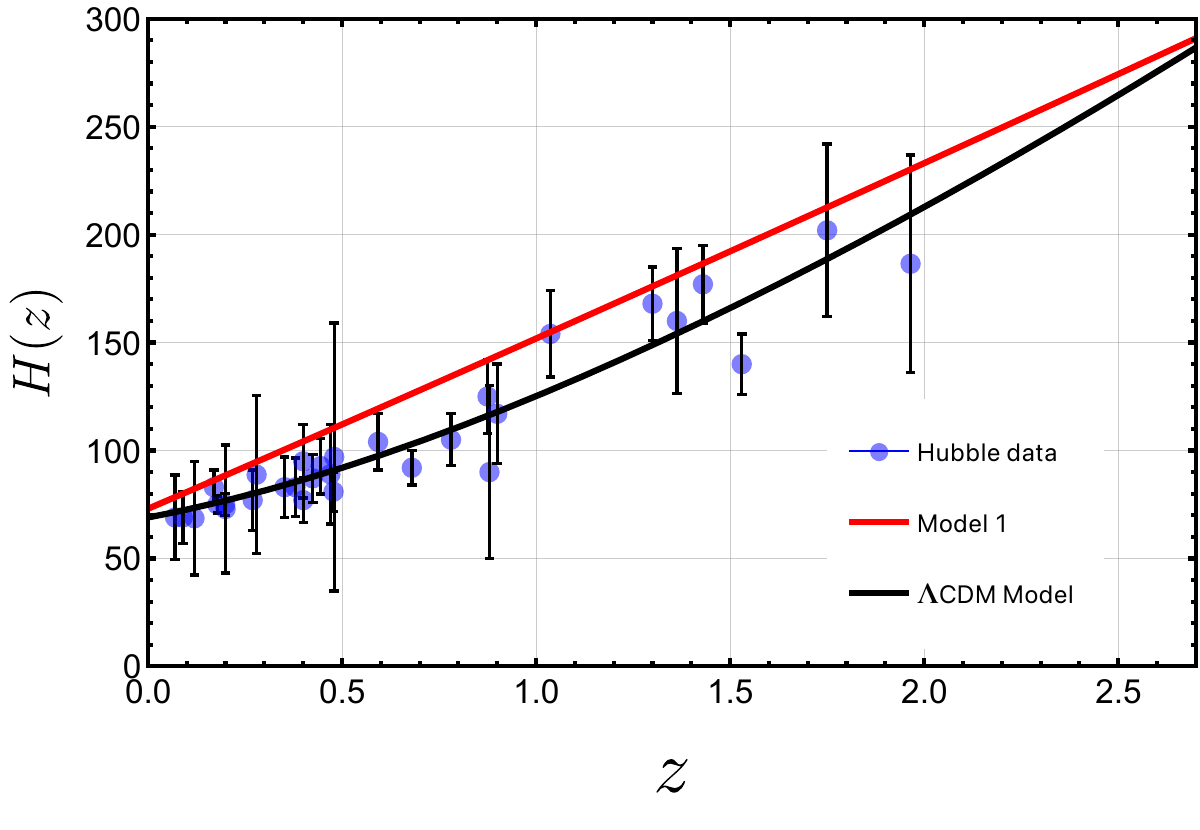}
    \caption{Plot of $H(z)$ parameter of Model II.}
    \label{fig_3b}
\end{subfigure}\hfil
\caption{The figure shows the theoretical curve of the Hubble parameter $H(z)$ for \(\Lambda\)CDM (black line), along with Model 1 (red line) and Model 2 (purple line), compared against the observable Hubble data (blue points with error bars in black line). For the \(\Lambda\)CDM model, the parameters were fixed at \(\Omega_{\mathrm{m0}} = 0.3\) and \(\Omega_\Lambda = 0.7\).}\label{fig_4}
\end{figure*}
\section{Information Criteria}
To analyze the sustainability of our model, we need to use one of the Information Criteria (IC), such as the Akaike Information Criterion (AIC) \cite{AIC1}, which is commonly used in model selection and is given by \cite{AIC2,AIC3,AIC4,AIC5}: $\text{AIC} = -2\ln (\mathcal{L}_{max}) + 2\kappa + \frac{2\kappa(\kappa + 1)}{N - \kappa - 1},$ where \( \mathcal{L}_{max} \) is the maximum likelihood function, \( \kappa \) is the total number of model parameters, and \( N \) is the total number of data points used to fit the model. For large \( N \), this formula reduces to the simpler expression: $\text{AIC} = -2\ln (\mathcal{L}_{max}) + 2\kappa.$
The AIC can be used to compare the goodness-of-fit of different models, with the model having the lowest AIC being the most favored. To compare models, we can compute the differences \( \Delta \text{AIC} \) and \( \Delta \text{BIC} \) relative to a $\Lambda$CDM model. These differences are defined as: $\Delta \text{AIC} = \text{AIC}_{\text{model}} - \text{AIC}_{\text{$\Lambda$CDM}},$ and $\Delta \text{BIC} = \text{BIC}_{\text{model}} - \text{BIC}_{\text{$\Lambda$CDM}}.$ A negative value of \( \Delta \text{AIC} \) and \( \Delta \text{BIC} \) indicates that the model is preferred over the $\Lambda$CDM Model. If \( 0 < |\Delta \text{AIC}| \leq 2 \), the models are considered comparable, while \( |\Delta \text{AIC}| \geq 4 \) suggests that the model with the higher AIC is significantly less favored. For \( \Delta \text{BIC} \), similar rules apply: if \( 0 < |\Delta \text{BIC}| \leq 2 \), the disfavor is weak; \( 2 < |\Delta \text{BIC}| \leq 6 \) indicates strong disfavor, and \( |\Delta \text{BIC}| > 6 \) points to very strong disfavor. Additionally, we can assess the overall fit quality of our model using the reduced chi-squared, defined as: $\chi_{\text{red}}^2 = \frac{\chi_{\text{tot},min}^2}{N - \kappa},$ where \( \chi_{\text{tot},min}^2 \) is the minimized total chi-squared value. The reduced chi-squared helps determine how well the model fits the data: A value of \( \chi_{\text{red}}^2 \approx 1 \) indicates a good fit, suggesting that the model adequately describes the data within expected statistical fluctuations. If \( \chi_{\text{red}}^2 < 1 \), it may indicate overfitting, where the model is too complex for the data. Conversely, if \( \chi_{\text{red}}^2 \gg 1 \), this suggests a poor fit, indicating that the model fails to capture important features of the data. By utilizing AIC, \( \Delta \text{AIC} \), \( \Delta \text{BIC} \), and \( \chi_{\text{red}}^2 \), we can comprehensively evaluate the sustainability and performance of our model in relation to simpler $\Lambda$CDM models.\\\\
\begin{table*}
\centering
\begin{tabular}{c c c c c c c c}
\hline
Model & $\chi_{min}^{2}$ & $\kappa$ & $\chi_{red}^{2}$ & $AIC$ & $\Delta AIC$ & $BIC$ & $\Delta BIC$ \\[1ex] \hline \hline
$\Lambda$CDM Model & 1036.47 & 3 & 0.991 & 1042.47 & 0 & 1057.33 & 0 \\ 
Model I & 1027.59 & 5 & 0.983 & 1037.59 & -4.88 & 1062.36 & 5.03 \\
Model II & 1024.46  & 6  & 0.982 & 1036.46 & -6.01 & 1066.34 & 9.01  \\
\hline\hline
\end{tabular}
\caption{Summary of the ${{\protect\chi}^2_{min}}$, ${{\protect\chi}^2_{red}}
$, $AIC$, $\Delta AIC$, $BIC$ and $\Delta BIC$.}
\label{table_2}
\end{table*}
\begin{table*}
\centering
\begin{tabular}{c c c c c c c}
\hline
Model & $q(z=0)$ & $q(z_t)$ & $w(z=0)$ & $Om(z=0)$ & $\Omega_{m}(z=0)$ & $\Omega_{eff}(z=0)$ \\[1ex] \hline \hline
   $\Lambda$CDM Model & $-0.535$ & $0.677$ & $-1$ & $0.30$ & $0.30$ & $0.70$ \\ 
Model I & $-0.343$ & --- & $-1.45$ & $0.438$ & $0.321$ & $0.679$  \\ 
Model II & $-0.607$ &$0.577$ & $-0.114$ & $0.267$ & $0.266$ & $0.734$ \\ \hline\hline
\end{tabular}%
\caption{ The qualitative behavior of the models under consideration along with the $\Lambda$CDM model.}
\label{table_3}
\end{table*}
\section{$Om$ diagnostics}
\label{section 5}
In this study, we present a different approach that allows us to distinguish $\Lambda$CDM from other DE models without using the cosmic EOS directly  \cite{54}. The $Om$ is geometrical diagnostics, and it is defined with the help of the Hubble parameter and redshift as
\begin{equation}
    Om(z)=\frac{\left(\frac{H(z)}{H_0}\right)^2-1}{(1+z)^3-1},
\end{equation}
where $H_0$ is the Hubble constant.\\
 By observing the slope of $Om(z)$ trajectory, this diagnostic can be used to discriminate between various Dark energy models. The Dark energy behaves like a quintessence if the slope of $Om(z)$ trajectory is negative, whereas if the slope of $Om(z)$ trajectory is positive, the Dark energy behaves like a phantom. The zero slope of $Om(z)$ trajectory (means constant behavior of $Om(z)$) represents that Dark energy is a cosmological constant ($\Lambda$CDM).
\begin{figure}
\includegraphics[scale=0.65]{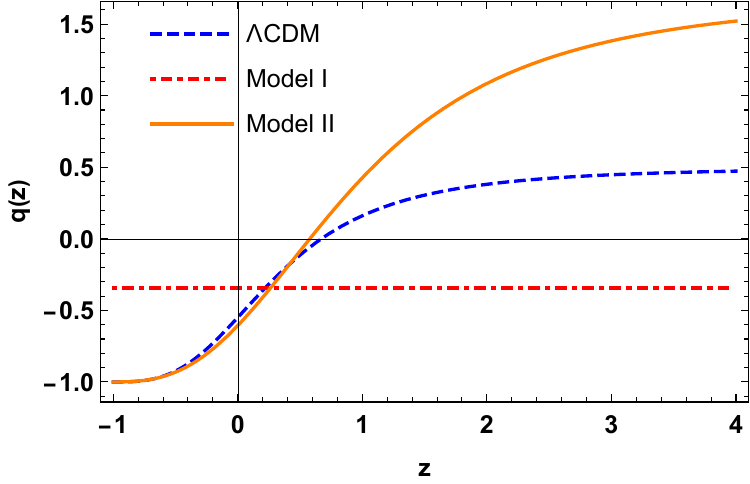}
\caption{The evolutionary trajectory of $q(z)$ versus redshift $z$.}
\label{q}
\end{figure}
\begin{figure}
\includegraphics[scale=0.65]{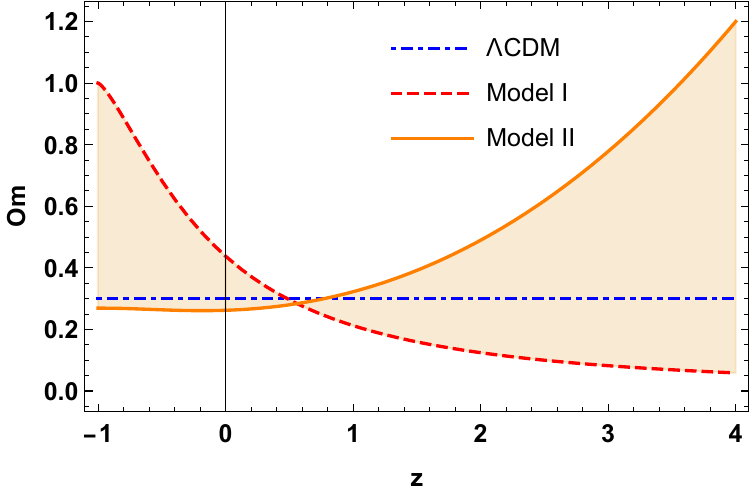}
\caption{The evolutionary trajectory of $Om(z)$ versus redshift $z$.}
\label{om}
\end{figure}
\section{Conclusion}\label{section 6}
Explaining late-time acceleration is a major challenge in cosmology. The Weyl-type $f(Q,T)$ gravity theory has been proposed to address this. In this theory, the non-metricity $Q_{\sigma\mu\nu}$ of space-time is represented in its standard Weyl form and is defined by a vector field $\omega_{\mu}$.\\
We used the $SNeIa$ $Pantheon+$ sample to constrain the free parameters of the Weyl-type \( f(Q, T) \) gravity framework. Figs ~\ref{fig_1} and ~\ref{fig_2} display the constraints on the free parameters of Model I and Model II. The \(1\sigma\) confidence interval indicates the region where approximately 68\% of the likelihood is concentrated, and the \(2\sigma\) confidence interval encompasses about 95\% of the likelihood distribution. Table \ref{table_1} presents the constraints at the 95\% confidence level (CL) on the cosmological parameters of the standard \(\Lambda\)CDM model, as well as Models I and II. We also demonstrate the compatibility of Models I and II with the $\Lambda$CDM model, using the Pantheon$^+$ and Cosmic Chronometers (CC) datasets. Fig~\ref{fig_1} shows the evolution of $\mu(z)$, indicating that both models display strong agreement with the $\Lambda$CDM model and the Pantheon$^+$ data. Fig~\ref{fig_2} presents the evolution of the Hubble parameter $H(z)$, where both models exhibit a slight deviation. However, this deviation is minimal, and both models still predict a similar evolutionary history as the $\Lambda$CDM model. The \(\Lambda\)CDM model is widely accepted, we compared our models with it to understand the differences in detail. The comparison utilized the Akaike Information Criteria (AIC), which considers both the goodness of fit (measured by \(\chi^2_{\text{min}}\)) and model complexity (number of parameters). Table \ref{table_2} show the values of \(\chi^2_{\text{min}}\), \(\chi^2_{\text{red}}\), AIC, \(\Delta\)AIC, BIC, and \(\Delta\)BIC. Our analysis indicated that the considered models are generally consistent with the \(\Lambda\)CDM model.\\\\
Cosmological parameters, such as the equation-of-state (EoS) and deceleration parameters, are vital for understanding the Universe's evolution. The EoS parameter, in particular, helps describe different fluid-like behaviors of the Universe. For instance, it can indicate a quintessence scenario ($-1<w<-1/3$) or a phantom scenario ($w<-1$). In this study, model I predicts the dark energy (DE) EoS parameter as $w=-1.096_{-0.065}^{+0.099}$ and the matter density parameter as $\Omega_{m,0}=0.321$. Model II predicts these values as $w=-0.153_{-0.087}^{+0.175}$ and $\Omega_{m,0}=0.266$. The values from model I align with the 2018 Planck data predictions \cite{collaboration2020planck}.
Fig ~\ref{q} represents the behavior of the deceleration parameter ($q$) for both models. At high redshift, in the case of Model I, a discrepancy is observed between its numerical value and the standard $\Lambda$CDM  paradigm. This suggests that in Model I, the Universe experiences slower expansion. As the redshift decreases, the discrepancy between the numerical values of Model I and the standard $\Lambda$CDM model also decreases, leading to a closer alignment between Model I and the standard $\Lambda$CDM  paradigm. On the other hand, Model II exhibits a linear behavior throughout its evolution, maintaining a constant value of $q_0=-0.343$ at both high and low redshifts. Further, we tested our model using the $Om$ diagnostic and compared it to the $\Lambda$CDM model to assess its dark energy profile. Fig ~\ref{om} shows the evolution of $Om(z)$ with redshift ($z$) for each model. Notably, In the case of Model I, We observe that for redshift values exceeding 0.3, the $Om(z)$ remains consistently below the current matter density parameter, denoted as $\Omega_{m0}$. This behavior suggests that the model predominantly falls within the phantom region during this redshift range. However, as the Universe continues to evolve, the $Om(z)$ begins to surpass the current matter density parameter $\Omega_{m0}$, indicating a transition into the quintessence domain. In the case of Model II, within the redshift range where ($z>$ 0.6), the value of $Om(z)$ exceeds the current matter density parameter $\Omega_{m0}$. This suggests that Model 2 falls within the quintessence domain. However, as the redshift decreases ( $z<$ 0.6), the value of $Om(z)$ becomes lower than the current matter density parameter $\Omega_{m0}$, indicating that the model enters the Phantom region. \\
Finally, Weyl type \( f(Q,T) \) models demonstrate strong alignment with observational data and present a compelling candidate for describing the nature of dark energy. Both linear and nonlinear variants exhibit robust compatibility with the \(\Lambda\)CDM model from statistical and cosmological perspectives. However, the theoretical viability of these models hinges on their ability to maintain mathematical consistency, including the absence of singularities or instabilities in their field equations, and ensuring adherence to general covariance principles. Observationally, they must continue to meet stringent tests from cosmological observations, solar system dynamics, and emerging gravitational wave studies to establish their credibility as viable alternatives to General Relativity and \( f(R) \) gravity. Compared to \( f(R) \) gravity, which has been extensively studied and tightly constrained by data, Weyl \( f(Q,T) \) models offer more flexibility through the additional degrees of freedom introduced by \( Q \) and \( T \). However, this increased complexity also presents challenges in fitting observational data and avoiding theoretical issues like ghost instabilities or strong coupling problems.\\
In conclusion, while this study primarily focuses on the observational aspects of Weyl $f(Q,T)$ gravity, the evaluation of perturbations remains a crucial area for future research. Analyzing perturbations can provide insights into the stability of the theory, the propagation of gravitational waves, and the growth of cosmic structures. These aspects are essential for comparing theoretical predictions with observational data, thus enhancing our understanding of Weyl $f(Q,T)$ gravity and its implications. Future work should aim to thoroughly investigate these perturbative aspects to further test and constrain the theory.\\\\
\textbf{Data availability} There are no new data associated with this article.
\section{Acknowledgements}
GNG acknowledges University Grants Commission (UGC), New Delhi, India for awarding Junior Research Fellowship (UGC-Ref. No.: 201610122060). PKS acknowledges the Science and Engineering Research Board, Department of Science and Technology, Government of India for financial support to carry out Research Project No.: CRG/2022/001847 and IUCAA, Pune, India for providing support through the visiting Associateship program. 
We are very much grateful to the honorable referees and to the editor for the illuminating suggestions that have significantly improved our work in terms of research quality, and presentation.\\\\

\bibliographystyle{elsarticle-num}
\bibliography{mybib}

\begin{thebibliography}{10}
\expandafter\ifx\csname url\endcsname\relax
  \def\url#1{\texttt{#1}}\fi
\expandafter\ifx\csname urlprefix\endcsname\relax\def\urlprefix{URL }\fi
\expandafter\ifx\csname href\endcsname\relax
  \def\href#1#2{#2} \def\path#1{#1}\fi

\bibitem{1:2001qse}
A.~G. Riess, et~al., {The farthest known supernova: support for an accelerating
  universe and a glimpse of the epoch of deceleration}, Astrophys. J. 560
  (2001) 49--71.
\newblock \href {http://arxiv.org/abs/astro-ph/0104455}
  {\path{arXiv:astro-ph/0104455}}, \href {https://doi.org/10.1086/322348}
  {\path{doi:10.1086/322348}}.

\bibitem{2}
S.~Perlmutter, M.~S. Turner, M.~White, Constraining dark energy with type ia
  supernovae and large-scale structure, Physical Review Letters 83~(4) (1999)
  670.

\bibitem{3}
{First year Wilkinson Microwave Anisotropy Probe (WMAP) observations:
  Preliminary maps and basic results}, Astrophys. J. Suppl. 148 (2003) 1--27.
\newblock \href {http://arxiv.org/abs/astro-ph/0302207}
  {\path{arXiv:astro-ph/0302207}}, \href {https://doi.org/10.1086/377253}
  {\path{doi:10.1086/377253}}.

\bibitem{4}
{First year Wilkinson Microwave Anisotropy Probe (WMAP) observations: The
  Angular power spectrum}, Astrophys. J. Suppl. 148 (2003) 135.
\newblock \href {http://arxiv.org/abs/astro-ph/0302217}
  {\path{arXiv:astro-ph/0302217}}, \href {https://doi.org/10.1086/377225}
  {\path{doi:10.1086/377225}}.

\bibitem{5}
D.~J. Eisenstein, I.~Zehavi, D.~W. Hogg, R.~Scoccimarro, M.~R. Blanton, R.~C.
  Nichol, R.~Scranton, H.-J. Seo, M.~Tegmark, Z.~Zheng, et~al., Detection of
  the baryon acoustic peak in the large-scale correlation function of sdss
  luminous red galaxies, The Astrophysical Journal 633~(2) (2005) 560.

\bibitem{6}
T.~Koivisto, D.~F. Mota, Dark energy anisotropic stress and large scale
  structure formation, Physical Review D 73~(8) (2006) 083502.

\bibitem{7}
P.~J.~E. Peebles, B.~Ratra, The cosmological constant and dark energy, Reviews
  of modern physics 75~(2) (2003) 559.

\bibitem{8}
V.~Sahni, A.~A. Starobinsky, {The Case for a positive cosmological Lambda
  term}, Int. J. Mod. Phys. D 9 (2000) 373--444.
\newblock \href {http://arxiv.org/abs/astro-ph/9904398}
  {\path{arXiv:astro-ph/9904398}}, \href
  {https://doi.org/10.1142/S0218271800000542}
  {\path{doi:10.1142/S0218271800000542}}.

\bibitem{9}
H.~A. Buchdahl, Non-linear lagrangians and cosmological theory, Monthly Notices
  of the Royal Astronomical Society 150~(1) (1970) 1--8.

\bibitem{10}
A.~A. Starobinsky, Disappearing cosmological constant in $f(r)$ gravity, JETP
  letters 86 (2007) 157--163.

\bibitem{clifton2012modified}
T.~Clifton, P.~G. Ferreira, A.~Padilla, C.~Skordis, Modified gravity and
  cosmology, Physics reports 513~(1-3) (2012) 1--189.

\bibitem{nojiri2011unified}
S.~Nojiri, S.~D. Odintsov, Unified cosmic history in modified gravity: from f
  (r) theory to lorentz non-invariant models, Physics Reports 505~(2-4) (2011)
  59--144.

\bibitem{nojiri2017modified}
S.~Nojiri, S.~Odintsov, V.~Oikonomou, Modified gravity theories on a nutshell:
  Inflation, bounce and late-time evolution, Physics Reports 692 (2017) 1--104.

\bibitem{Odintsov:2023weg}
S.~D. Odintsov, V.~K. Oikonomou, I.~Giannakoudi, F.~P. Fronimos, E.~C.
  Lymperiadou, {Recent Advances on Inflation} (7 2023).
\newblock \href {http://arxiv.org/abs/2307.16308} {\path{arXiv:2307.16308}}.

\bibitem{11}
S.~Capozziello, V.~Cardone, H.~Farajollahi, A.~Ravanpak, Cosmography in $f(t)$
  gravity, Physical Review D 84~(4) (2011) 043527.

\bibitem{12}
T.~Harko, F.~S. Lobo, S.~Nojiri, S.~D. Odintsov, f (r, t) gravity, Physical
  Review D 84~(2) (2011) 024020.

\bibitem{harko2014f}
T.~Harko, F.~S. Lobo, G.~Otalora, E.~N. Saridakis, $f(\mathcal{T},t)$ gravity
  and cosmology, Journal of Cosmology and Astroparticle Physics 2014~(12)
  (2014) 021.

\bibitem{13}
N.~Goheer, R.~Goswami, P.~K. Dunsby, K.~Ananda, Coexistence of matter dominated
  and accelerating solutions in f (g) gravity, Physical Review D 79~(12) (2009)
  121301.

\bibitem{14}
E.~Elizalde, R.~Myrzakulov, V.~V. Obukhov, D.~Saez-Gomez, {$\Lambda$CDM epoch
  reconstruction from F(R,G) and modified Gauss-Bonnet gravities}, Class.
  Quant. Grav. 27 (2010) 095007.
\newblock \href {http://arxiv.org/abs/1001.3636} {\path{arXiv:1001.3636}},
  \href {https://doi.org/10.1088/0264-9381/27/9/095007}
  {\path{doi:10.1088/0264-9381/27/9/095007}}.

\bibitem{15}
J.~M. Nester, H.-J. Yo, Symmetric teleparallel general relativity, arXiv
  preprint gr-qc/9809049 (1998).

\bibitem{16}
J.~B. Jim{\'e}nez, L.~Heisenberg, T.~Koivisto, Coincident general relativity,
  Physical Review D 98~(4) (2018) 044048.

\bibitem{17}
T.~Harko, T.~S. Koivisto, F.~S. Lobo, G.~J. Olmo, D.~Rubiera-Garcia, Coupling
  matter in modified q gravity, Physical Review D 98~(8) (2018) 084043.

\bibitem{18}
S.~Mandal, D.~Wang, P.~Sahoo, Cosmography in f(q) gravity, Physical Review D
  102~(12) (2020) 124029.

\bibitem{19}
N.~Frusciante, Signatures of f(q) gravity in cosmology, Physical Review D
  103~(4) (2021) 044021.

\bibitem{20}
W.~Khyllep, A.~Paliathanasis, J.~Dutta, Cosmological solutions and growth index
  of matter perturbations in f(q) gravity, Physical Review D 103~(10) (2021)
  103521.

\bibitem{22}
G.~N. Gadbail, S.~Mandal, P.~Sahoo, Reconstruction of $\lambda$cdm universe in
  f (q) gravity, Physics Letters B 835 (2022) 137509.

\bibitem{23}
S.~Capozziello, R.~D'Agostino, Model-independent reconstruction of f(q)
  non-metric gravity, Physics Letters B 832 (2022) 137229.

\bibitem{24}
W.~Wang, H.~Chen, T.~Katsuragawa, Static and spherically symmetric solutions in
  f (q) gravity, Physical Review D 105~(2) (2022) 024060.

\bibitem{Gadbail_APJ_2024}
G.~N. Gadbail, S.~Mandal, P.~K. Sahoo, Gaussian process approach for
  model-independent reconstruction of $f(q)$ gravity with direct hubble
  measurements, The Astrophysical Journal 972~(2) (2024) 174.
\newblock \href {https://doi.org/10.3847/1538-4357/ad5cf4}
  {\path{doi:10.3847/1538-4357/ad5cf4}}.

\bibitem{21}
I.~Ayuso, R.~Lazkoz, V.~Salzano, Observational constraints on cosmological
  solutions of f(q) theories, Physical review d 103~(6) (2021) 063505.

\bibitem{anagnostopoulos2021first}
F.~K. Anagnostopoulos, S.~Basilakos, E.~N. Saridakis, First evidence that
  non-metricity f (q) gravity could challenge $\lambda$cdm, Physics Letters B
  822 (2021) 136634.

\bibitem{anagnostopoulos2023new}
F.~K. Anagnostopoulos, V.~Gakis, E.~N. Saridakis, S.~Basilakos, New models and
  big bang nucleosynthesis constraints in f (q) gravity, The European Physical
  Journal C 83~(1) (2023) 58.

\bibitem{25}
Y.~Xu, G.~Li, T.~Harko, S.-D. Liang, f (q, t) gravity, The European Physical
  Journal C 79 (2019) 1--19.

\bibitem{26:2020tuk}
S.~Arora, S.~K.~J. Pacif, S.~Bhattacharjee, P.~K. Sahoo, {$f(Q,T)$ gravity
  models with observational constraints}, Phys. Dark Univ. 30 (2020) 100664.
\newblock \href {http://arxiv.org/abs/2007.01703} {\path{arXiv:2007.01703}},
  \href {https://doi.org/10.1016/j.dark.2020.100664}
  {\path{doi:10.1016/j.dark.2020.100664}}.

\bibitem{27}
S.~Arora, A.~Parida, P.~Sahoo, Constraining effective equation of state in f
  (q, t) gravity, The European Physical Journal C 81 (2021) 1--7.

\bibitem{28}
G.~N. Gadbail, S.~Arora, P.~Sahoo, Generalized chaplygin gas and accelerating
  universe in f (q, t) gravity, Physics of the Dark Universe 37 (2022) 101074.

\bibitem{Arora_2021E}
S.~Arora, J.~Santos, P.~Sahoo,
  \href{http://dx.doi.org/10.1016/j.dark.2021.100790}{Constraining f(q,t)
  gravity from energy conditions}, Physics of the Dark Universe 31 (2021)
  100790.
\newblock \href {https://doi.org/10.1016/j.dark.2021.100790}
  {\path{doi:10.1016/j.dark.2021.100790}}.
\newline\urlprefix\url{http://dx.doi.org/10.1016/j.dark.2021.100790}

\bibitem{Arora_2020E}
S.~Arora, P.~K. Sahoo, \href{http://dx.doi.org/10.1088/1402-4896/abaddc}{Energy
  conditions in f(q, t) gravity}, Physica Scripta 95~(9) (2020) 095003.
\newblock \href {https://doi.org/10.1088/1402-4896/abaddc}
  {\path{doi:10.1088/1402-4896/abaddc}}.
\newline\urlprefix\url{http://dx.doi.org/10.1088/1402-4896/abaddc}

\bibitem{29}
S.~Bhattacharjee, P.~Sahoo, Baryogenesis in f (q, t) gravity, The European
  Physical Journal C 80~(3) (2020) 289.

\bibitem{30}
A.~N{\'a}jera, A.~Fajardo, Cosmological perturbation theory in f (q, t)
  gravity, Journal of Cosmology and Astroparticle Physics 2022~(03) (2022) 020.

\bibitem{31}
G.~N. Gadbail, S.~Arora, P.~Sahoo, Reconstruction of f (q, t) lagrangian for
  various cosmological scenario, Physics Letters B (2023) 137710.

\bibitem{Loo_2023}
T.-H. Loo, R.~Solanki, A.~De, P.~K. Sahoo,
  \href{http://dx.doi.org/10.1140/epjc/s10052-023-11391-4}{f(q,t) gravity, its
  covariant formulation, energy conservation and phase-space analysis}, The
  European Physical Journal C 83~(3) (Mar. 2023).
\newblock \href {https://doi.org/10.1140/epjc/s10052-023-11391-4}
  {\path{doi:10.1140/epjc/s10052-023-11391-4}}.
\newline\urlprefix\url{http://dx.doi.org/10.1140/epjc/s10052-023-11391-4}

\bibitem{32}
M.~Tayde, Z.~Hassan, P.~Sahoo, S.~Gutti, Static spherically symmetric wormholes
  in gravity, Chinese Physics C 46~(11) (2022) 115101.

\bibitem{33}
M.~Tayde, S.~Ghosh, P.~K. SAHOO, Non-exotic static spherically symmetric
  thin-shell wormhole solution in f(q,t) gravity, Chinese Physics C (2023).

\bibitem{34}
Y.~Xu, T.~Harko, S.~Shahidi, S.-D. Liang, Weyl type f (q, t) gravity, and its
  cosmological implications, The European Physical Journal C 80~(5) (2020)
  1--22.

\bibitem{35}
J.-Z. Yang, S.~Shahidi, T.~Harko, S.-D. Liang, Geodesic deviation, raychaudhuri
  equation, newtonian limit, and tidal forces in weyl-type f (q, t) gravity,
  The European Physical Journal C 81 (2021) 1--19.

\bibitem{36}
G.~Gadbail, S.~Arora, P.~Sahoo, Power-law cosmology in weyl-type f (q, t)
  gravity, The European Physical Journal Plus 136~(10) (2021) 1040.

\bibitem{37}
G.~N. Gadbail, S.~Arora, P.~Sahoo, Viscous cosmology in the weyl-type f (q, t)
  gravity, The European Physical Journal C 81~(12) (2021) 1088.

\bibitem{gadbail2022interaction}
G.~N. Gadbail, S.~Arora, P.~Kumar, P.~Sahoo, Interaction of divergence-free
  deceleration parameter in weyl-type f (q, t) gravity, Chinese Journal of
  Physics 79 (2022) 246--255.

\bibitem{gadbail2023dark}
G.~N. Gadbail, S.~Arora, P.~Sahoo, Dark energy constraint on equation of state
  parameter in the weyl type f (q, t) gravity, Annals of Physics 451 (2023)
  169244.

\bibitem{38}
H.~Weyl, Gravitation und elektrizit{\"a}t', sitzungsberichte der preu{\"y},
  Akad. Berlin, Math. Kl (1918) 465.

\bibitem{39}
{Observational constraints on Hubble constant and deceleration parameter in
  power-law cosmology}, Mon. Not. Roy. Astron. Soc. 422 (2012) 2532--2538.
\newblock \href {http://arxiv.org/abs/1109.6924} {\path{arXiv:1109.6924}},
  \href {https://doi.org/10.1111/j.1365-2966.2012.20810.x}
  {\path{doi:10.1111/j.1365-2966.2012.20810.x}}.

\bibitem{40}
J.~Singh, S.~Rani, Modified chaplygin gas cosmology with statefinder diagnostic
  in lyra geometry, Applied Mathematics and Computation 259 (2015) 187--197.

\bibitem{41}
C.~Zhang, H.~Zhang, S.~Yuan, S.~Liu, T.-J. Zhang, Y.-C. Sun, Four new
  observational h (z) data from luminous red galaxies in the sloan digital sky
  survey data release seven, Research in Astronomy and Astrophysics 14~(10)
  (2014) 1221.

\bibitem{42}
D.~M. Scolnic, D.~Jones, A.~Rest, Y.~Pan, R.~Chornock, R.~Foley, M.~Huber,
  R.~Kessler, G.~Narayan, A.~Riess, et~al., The complete light-curve sample of
  spectroscopically confirmed sne ia from pan-starrs1 and cosmological
  constraints from the combined pantheon sample, The Astrophysical Journal
  859~(2) (2018) 101.

\bibitem{43}
D.~Brout, D.~Scolnic, B.~Popovic, A.~G. Riess, A.~Carr, J.~Zuntz, R.~Kessler,
  T.~M. Davis, S.~Hinton, D.~Jones, et~al., The pantheon+ analysis:
  cosmological constraints, The Astrophysical Journal 938~(2) (2022) 110.

\bibitem{44}
D.~Scolnic, D.~Brout, A.~Carr, A.~G. Riess, T.~M. Davis, A.~Dwomoh, D.~O.
  Jones, N.~Ali, P.~Charvu, R.~Chen, et~al., The pantheon+ analysis: The full
  data set and light-curve release, The Astrophysical Journal 938~(2) (2022)
  113.

\bibitem{48}
R.~Kessler, D.~Scolnic, Correcting type ia supernova distances for selection
  biases and contamination in photometrically identified samples, The
  Astrophysical Journal 836~(1) (2017) 56.

\bibitem{AIC1}
H.~Akaike, A new look at the statistical model identification, IEEE
  transactions on automatic control 19~(6) (1974) 716--723.

\bibitem{AIC2}
M.~Li, X.~Li, X.~Zhang, Comparison of dark energy models: A perspective from
  the latest observational data, Science China Physics, Mechanics and Astronomy
  53 (2010) 1631--1645.

\bibitem{AIC3}
K.~P. Burnham, D.~R. Anderson, K.~P. Huyvaert, Aic model selection and
  multimodel inference in behavioral ecology: some background, observations,
  and comparisons, Behavioral ecology and sociobiology 65 (2011) 23--35.

\bibitem{AIC4}
K.~P. Burnham, D.~R. Anderson, Multimodel inference: understanding aic and bic
  in model selection, Sociological methods \& research 33~(2) (2004) 261--304.

\bibitem{AIC5}
A.~R. Liddle, Information criteria for astrophysical model selection, Monthly
  Notices of the Royal Astronomical Society: Letters 377~(1) (2007) L74--L78.

\bibitem{54}
V.~Sahni, A.~Shafieloo, A.~A. Starobinsky, Two new diagnostics of dark energy,
  Physical Review D 78~(10) (2008) 103502.

\bibitem{collaboration2020planck}
P.~Collaboration, N.~Aghanim, Y.~Akrami, M.~Ashdown, J.~Aumont, C.~Baccigalupi,
  M.~Ballardini, A.~Banday, R.~Barreiro, N.~Bartolo, et~al., Planck 2018
  results. vi. cosmological parameters (2020).

\end{thebibliography}

\end{document}